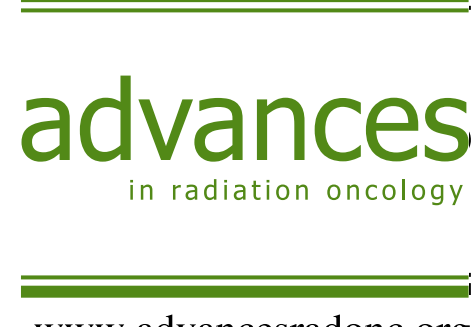

Scientific Article

# Cherenkov Imaged Bio-Morphological Features Verify Patient Positioning With Deformable Tissue Translocation in Breast Radiation Therapy

**Yao Chen, BE,[a] Savannah M. Decker, PhD,[a] Petr Bruza, PhD,[a] David J. Gladstone, ScD,[a,b,c] Lesley A. Jarvis, MD, PhD,[b,c] Brian W. Pogue, PhD,[a,b,c,d] Kimberley S. Samkoe, PhD,[a,b,c] and Rongxiao Zhang, PhD[a,c,e,*]**

[a]*Thayer School of Engineering, Dartmouth College, Hanover, New Hampshire;* [b]*Geisel School of Medicine, Dartmouth College, Hanover, New Hampshire;* [c]*Dartmouth Cancer Center, Dartmouth Health, Lebanon, New Hampshire;* [d]*Department of Medical Physics, University of Wisconsin-Madison, Madison, Wisconsin; and* [e]*Department of Radiation Oncology, University of Missouri-Columbia, Columbia, Missouri*

**Purpose:** Accurate patient positioning is crucial for precise radiation therapy dose delivery, as errors in positioning can profoundly influence treatment outcomes. This study introduces a novel application for loco-regional tissue deformation tracking via Cherenkov image analysis during fractionated breast cancer radiation therapy. The primary objective of this research was to develop and test an algorithmic method for Cherenkov-based position accuracy quantification, particularly for loco-regional deformations, which do not have an ideal method for quantification during radiation therapy.

**Methods and Materials:** Bio-morphological features in the Cherenkov images, such as vessels, were segmented. A rigid/nonrigid combined registration technique was employed to pinpoint both inter- and intrafractional positioning variations. The methodology was tested on an anthropomorphic chest phantom experiment via shifting a treatment couch with known distances and inducing respiratory motion to simulate interfraction setup uncertainties and intrafraction motions, respectively. It was then applied to a data set of fractionated whole breast radiation therapy human imaging (n = 10 patients).

**Results:** The methodology provided quantified positioning variations comprising 2 components: a global shift determined through rigid registration and a 2-dimensional variation map illustrating loco-regional tissue deformation quantified via nonrigid registration. Controlled phantom testing yielded an average accuracy of 0.83 mm for couch translations up to 20 mm in various directions. Analysis of clinical Cherenkov imaging data from 10 breast cancer patients compared with their first imaged fraction revealed an interfraction setup variation of $3.7 \pm 2.4$ mm in the global shift and loco-regional deformation up to $3.3 \pm 1.9$ mm (95th percentile of all regional deformation).

**Conclusions:** This study introduces the use of Cherenkov visualized bio-morphological features to quantify the global and local variations in patient positioning based on rigid and nonrigid registrations. This new approach demonstrates the feasibility of providing quantitative guidance for inter/intrafraction positioning, particularly for the loco-regional deformations that have been unappreciated in current practice with conventional imaging techniques.

Sources of support: This research was supported by the NCI cancer center support grant 5P30 CA023108-41, 5R44 CA265654-02, as well as support from NIH grant R01 EB023909.

Research data are stored in an institutional repository and will be shared upon request to the corresponding author.

*Corresponding author: Rongxiao Zhang, PhD; Email: rongxiao.zhang@health.missouri.edu

## Introduction

Breast cancer is the most common malignancy in women, with over 300,000 new cases estimated in 2023.[1] Radiation therapy (RT) has been established as the standard of care for early-stage breast cancer, alone or combined with surgery and chemotherapy, significantly improving local control and survival outcomes.[2] During RT, the precision of patient positioning is crucial to accurately target the tumor while sparing surrounding healthy tissue.[3] Excessive deviations in patient setup or treatment delivery can lead to negative consequences, including suboptimal treatment of the tumor, elevated healthy tissue toxicity, and secondary radiation-induced cancer risks.[4]

With technological advances, radiograph-based routine image guidance, including 2-dimensional (2D) x-ray imaging, 3-dimensional (3D) cone beam computed tomography, radiograph fluoroscopy, and megavoltage radiograph portal imaging, have been deployed in the standard of care for patient alignment. Optical surface-guided RT (SGRT) has been introduced and developed over the past 2 decades and is now widely adopted for patient setup verification and monitoring.[5] Adding to conventional SGRT techniques, Cherenkov imaging has emerged as a noncontact technique for real-time visualization of RT, as depicted in Fig. 1A.[6] Cherenkov light emission occurs when high-energy photon or electron beams interact with tissue, inducing light emission that resembles the radiation beam shape and superficial dose on the patient surface, as shown in Fig. 1B.[7,8] The current implementation of this technique uses intensified cameras that are time-gated to synchronize with the linear accelerator (LINAC) pulses, enabling the real-time capture of low-intensity Cherenkov light while effectively suppressing ambient room light.[9,10] This technology has paved the way for novel applications in RT, including recent studies on inter- and intrafractional verification of treatment delivery and error detection.[9,11]

The passive nature of Cherenkov imaging throughout the entire RT course means that this technique can be used to assess patient positioning accuracy and inform adjustments when needed.[5] Prior research has shown the efficacy of Cherenkov imaging in detecting incidents that were not identified by other pretreatment checks, highlighting its unique capabilities.[9] Further investigations have sought to quantify positioning variations using metrics such as the Dice similarity coefficient and mean distance to conformity on binarized images, as well as mutual information and the $\gamma$ passing rate on nonbinarized images, to better detect and quantify the magnitude of errors in patient positioning.[11,12]

Current technologies, including Cherenkov imaging, focus on verifying the global alignment based on rigid image registration.[13] It is well known that soft tissue deformations exist in the treatment region, although quantification of this remains challenging. While reviewing the Cherenkov images acquired from breast RT patients, it was noted that bio-morphological features such as the vasculature were enhanced because of larger optical absorption,[14,15] which can be observed in Fig. 1B. These bio-morphological features could be used as an intrinsic surrogate to quantify the loco-regional deformation. The advantages of using Cherenkov imaged bio-morphological features to quantify the positioning variations are illustrated in Fig. 1C; the differences in the bio-morphological features between 2 fractions, which could be clearly observed visually, reflect the existing setup variations that would likely be ignored in the global Cherenkov beam review. These findings inspire the utilization of Cherenkov imaged bio-morphological features to quantify variations in patient positions, particularly from loco-regional deformations that have been neglected or underestimated based on global image analysis.

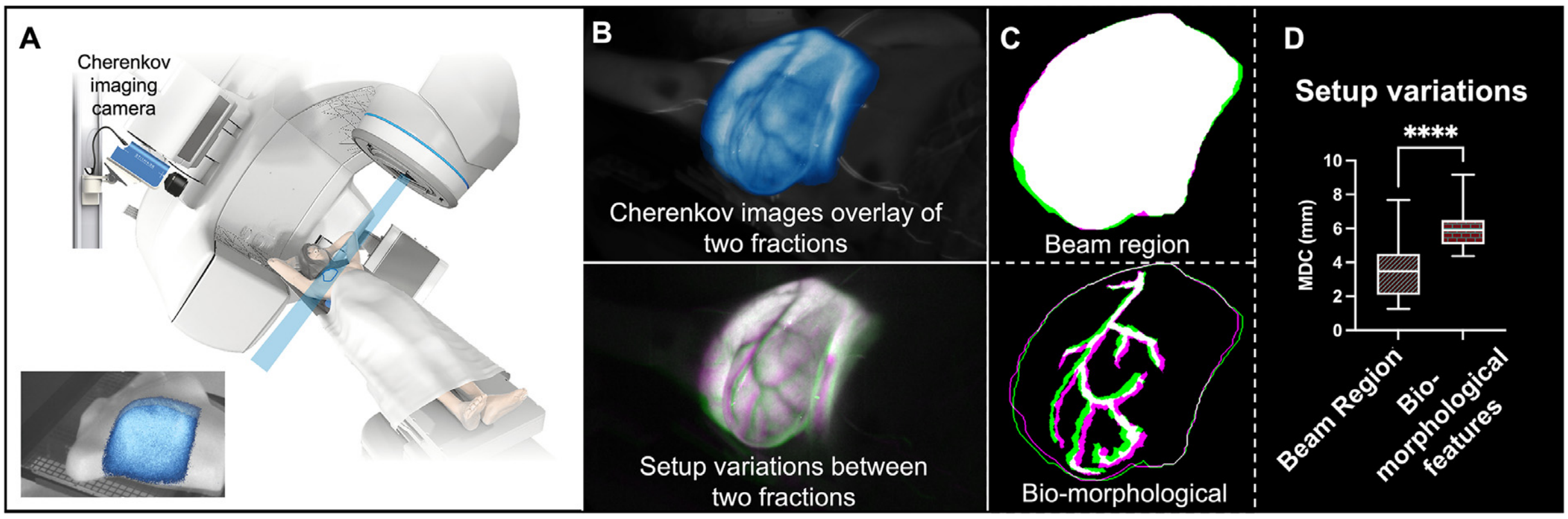

**Figure 1** The use of Cherenkov imaging for monitoring patient setup during radiation therapy for breast cancer. (A) Setup of Cherenkov imaging system during treatment. (B) Patient positioning variations captured by the Cherenkov imaging. The top panel shows 2 Cherenkov images from 2 fractions of the same patient, each 50% transparently overlaid onto the other to display the positioning variations. This 2-fraction overlaid composite image is then superimposed on the background white light image, with a pseudocolor map applied to Cherenkov images for enhanced visualization. The bottom panel displays the same pair of Cherenkov images but in grayscale and with 2 different color bands showing their setup variations existing in the 2 fractions, where the color of magenta represents the variation occurring in the moving image and green represents the reserved part in the reference image. (C) Comparison of observed setup variations between using the entire Cherenkov beam region and internal bio-morphological features. (D) Quantified setup variations using the entire Cherenkov beam regions and the internal bio-morphological features expressed in mean distance to conformity (MDC).

In this study, we explored using segmented bio-morphological features of the vasculature as internal biological markers in Cherenkov imaging to verify patient positioning accuracy and to inform future image guidance. We employed both rigid and nonrigid registration techniques on Cherenkov images acquired across different fractions (interfraction) or within a single fraction (intrafraction) to assess positioning accuracy, including the loco-regional deformations that have been neglected to date.

## Methods and Materials

### Clinical patient imaging

The patient images involved in this study were collected from a retrospective study approved by the institutional review board. Cherenkov cameras were mounted on the ceiling of the treatment room and were used to continuously capture images of Cherenkov emission while the radiation beam was on. Cherenkov images were displayed in real-time in the control room by BeamSite software (DoseOptics LLC). Per treatment fraction, the Cherenkov video frames were aggregated into a cumulative image that represented the entire video during that treatment session. Both the video footage and the cumulative image were preserved for subsequent analysis. This investigation analyzed Cherenkov data from 10 patients who had undergone RT for breast cancer listed in Table 1.

### Cherenkov image acquisition and processing

In this study, 2 Cherenkov imaging cameras were fixed to the ceiling of the treatment room and were angled toward the left and right sides of the treatment couch to capture the Cherenkov light emission from the patient's surface during radiation delivery (DoseOptics LLC). These cameras were synchronized and time-gated to the 4-microsecond pulses of the LINAC, allowing effective collection of Cherenkov signals during each pulse and rejecting room background signals in between pulses. To suppress the background noise in Cherenkov images, the background images were subtracted by the systems from the raw Cherenkov images. The background images were captured while the radiation beam was off immediately after each in-sync acquisition, and then they were processed by median filtering and dark field corrections to remove speckles and electronic noise, respectively. The median filter effectively removes speckle noise in the image by replacing each pixel's value with the median of neighboring pixel intensities, which eliminates outliers like speckle noise while preserving detailed edge information. Darkfield correction effectively reduces camera-related electronic noise by capturing a background image when the radiation beam is off, which subtracts the inherent sensor noise from the actual signal to improve the image quality. Details on the routine imaging settings and on-board processing have been described in previous publications.[16]

### Quantification methodology

More subtle setup variations have been observed based on the analysis using bio-morphological features compared with using the entire Cherenkov image, which is shown in Fig. 1D. A registration-based methodology using Cherenkov imaged bio-morphological features was

**Table 1 Patient demographics and breast radiation therapy treatment plans. The following information is provided for each patient: age at treatment, prescription dose, number of fractions imaged, and radiation therapy beam energy**

| Patient no. | Age | Sex | Site | Prescription dose | Beam energy |
|---|---|---|---|---|---|
| 1 | 63 | F | Left breast | 266 cGy × 16 fx | 6 + 10 MV |
| 2 | 69 | F | Left breast | 200 cGy × 25 fx | 6 + 10 MV |
| 3 | 69 | F | Left breast | 266 cGy × 16 fx | 6 + 10 MV |
| 4 | 51 | F | Left breast | 266 cGy × 16 fx | 6 MV |
| 5 | 68 | F | Left breast | 266 cGy × 16 fx | 6 + 10 MV |
| 6 | 58 | F | Right breast | 266 cGy × 16 fx | 6 MV |
| 7 | 67 | F | Right breast | 200 cGy × 25 fx | 6 MV |
| 8 | 66 | F | Right breast | 200 cGy × 25 fx | 6 MV |
| 9 | 70 | F | Right breast | 266 cGy × 16 fx | 6 MV |
| 10 | 73 | F | Right breast | 180 cGy × 25 fx | 6 + 10 MV |

*Abbreviation:* fx = fraction.

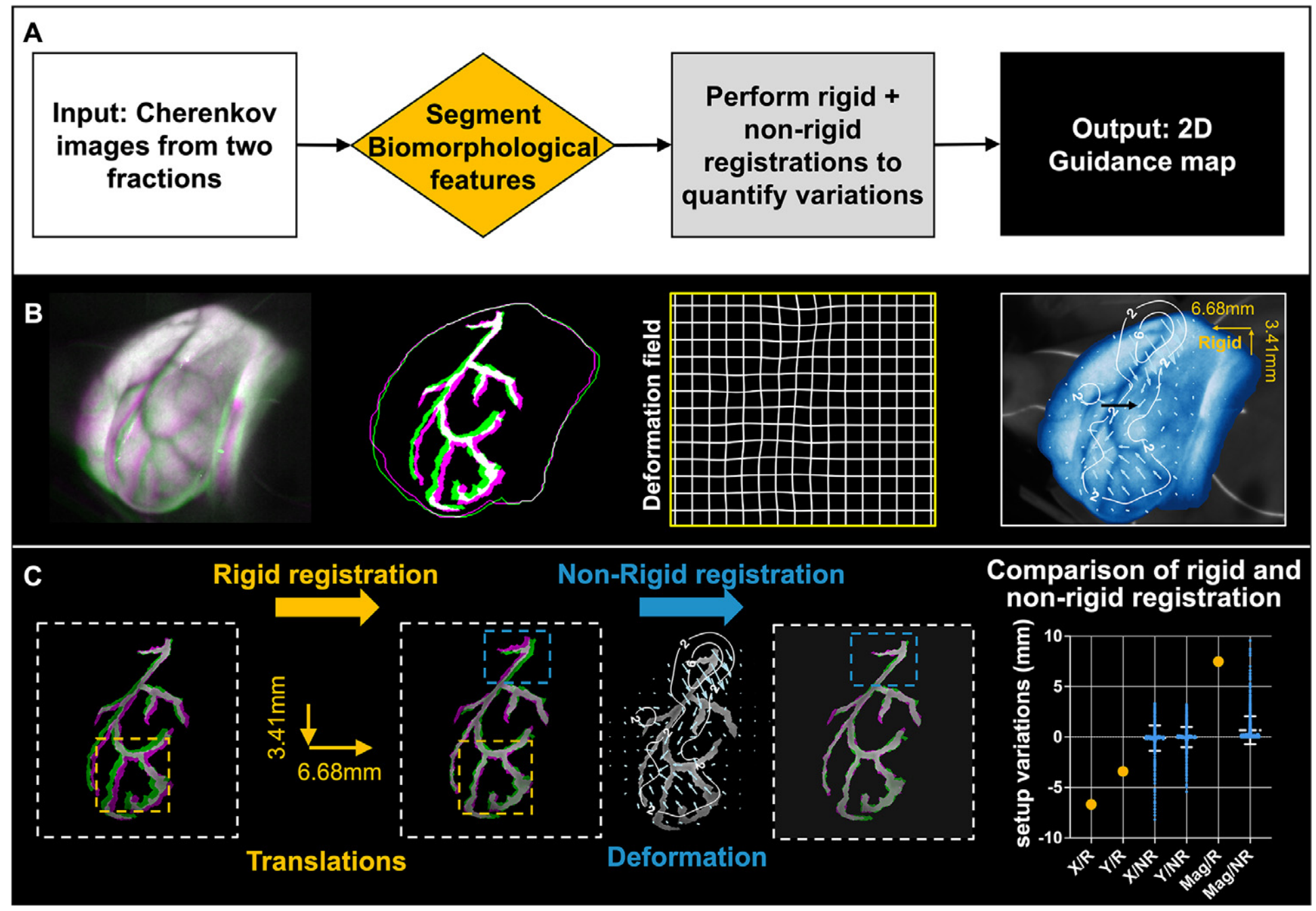

**Figure 2** Methodology to quantify patient positioning variations based on bio-morphological features in Cherenkov imaging. (A) Quantification workflow. (B) An example of patient Cherenkov image data undergoing this methodology to quantify interfraction setup variations. From left to right, each figure represents differences between the Cherenkov images of the moving fraction and the reference fraction; segmentation of bio-morphological features from 2 fractions and their differences; deformation field displayed in a grid map quantified from nonrigid registration; 2-dimensional (2D) correction guidance map: blue arrows overlaid on the Cherenkov image show the deformation vector for correcting the local deformation from the nonrigid registration, while orange arrows in the upper right corner indicate the global shifts in the *x* and *y* direction from the rigid registration. The white contour lines are curves along which the magnitude of the deformable correction vectors has a constant magnitude value. (C) Rigid and nonrigid registration combined workflow illustration. The orange arrows represent the translations from the first performed rigid registration, and the orange boxes highlight the area where rigid registration improves the alignment of bio-morphological features. The blue map represents the deformation in units of millimeters (mm) from the nonrigid registration performed sequentially, and the blue dotted boxes point out the area where nonrigid registration addresses the remaining loco-regional deformation. The rightmost box plot shows the comparison of variations quantified by the first performed rigid and sequentially performed nonrigid registration in terms of the translation in the *x* and *y* direction of the imaging plane and their magnitude. The global variations quantified by the rigid registration are represented by orange points in the box plot, whereas the pixel-level loco-regional deformation quantified by the nonrigid registration is displayed as a distribution in blue, where its mean and 1 SD shown in white are overlaid on its box plot.
*Abbreviations:* X/R = *X* direction by the Rigid reigistration; Y/NR = *Y* direction by the Non-Rigid registration; Mag = Magnitude.

developed in this study to quantify patient positioning variations. A flowchart is described in Fig. 2A, and each step is illustrated in Fig. 2B using a patient example. Image processing was completed in MATLAB version R2023b (MathWorks).

## Bio-morphological features segmentation

The first step of the quantification methodology is to segment the blood vessels. Because of large optical attenuation by blood in the wavelength of Cherenkov light,[17] clear contrast was observed between the vasculature and its surrounding tissues, which makes the superficial blood vessel a good candidate to be used as a bio-morphological feature. A contrast-enhanced filter was applied to the Cherenkov images, and manual adjustment of the contrast was employed to assist in the segmentation process. Multiplying the segmented binary vessel region of interest mask with the raw cumulative Cherenkov image, a bio-morphological feature image was generated for the later image registration step.

## Image registration

Image registration was applied to align the segmented bio-morphological features. A rigid and nonrigid combined registration scheme was deployed, as shown in Fig. 2C. A rigid registration was performed first to account for global shifts, followed by a nonrigid registration to account for the loco-regional deformation.

Rigid registration was performed using the *imregister* function in MATLAB, aiming to minimize the mean squared error between the bio-morphological features of 2 fractions, using a regular step gradient descent optimizer technique to solve the rigid image registration through translation and rotation.

Nonrigid registration was performed using a pixel-based B-spline grid-based registration algorithm.[18] The algorithm constructed a grid of basis spline control points that controlled the transformation of the input image and used the squared pixel distance as the similarity measure to assess the registration discrepancy between the reference and moving images. A fast optimization approach was applied using the limited-memory Broyden-Fletcher-Goldfarb-Shanno algorithm, a quasi-Newton method, which iteratively adjusted the control points to minimize registration errors until the optimal alignment between 2 images was achieved.[19]

With the combined scheme of rigid and nonrigid registration, both global shift and local deformation could be quantified from the 2 types of registration methods. In detail, a global translation indicating the shift in the $x$ and $y$ direction was extracted from the transformation matrix of the rigid registration result. A 2D map with the same size as the Cherenkov image indicating the pixel-level local tissue deformation was generated from the iterative process of nonrigid registration result. The magnitude of quantified variations from the registration-based methodology was in units of pixels but then converted to physical dimension with the unit of millimeter (mm) according to the 0.42 mm per pixel conversion factor from the Cherenkov acquisition system.

## Phantom study

Two phantom tasks were designed to test the accuracy of quantifying the interfraction and intrafraction variations. The steps are illustrated in Fig. 3A-D and E-G, respectively.

### Interfraction setup uncertainties

An anthropomorphic chest phantom, shown in Fig. 3A and designed to mimic the optical absorption properties of soft human tissue, was irradiated with a TrueBeam LINAC (Varian Medical Systems) using 6 MV x-rays. The irradiation followed an isocentric left breast treatment plan, which had been developed in Eclipse (Varian Medical Systems). The plan featured 2 opposing tangent beams aimed from the right anterior oblique (310°) and the left posterior oblique (135°) angles, delivering a dose of 266 cGy per treatment. A 2D pattern representing patient blood vessels was cut from a sheet of neutral density filter and placed flush against the phantom breast surface in the treatment region.

The interfraction setup uncertainties were simulated by the couch movement. The phantom was positioned according to the treatment plan and irradiated to capture a ground truth reference Cherenkov image, as shown in Fig. 3B. To create the variations among fractions and quantify the variations by the methodology, the treatment couch was systematically shifted in 5, 10, and 20 mm increments laterally, vertically, and longitudinally from the original treatment position. The same complete treatment was delivered at each displacement. A cumulative Cherenkov image was recorded at each displacement, resulting in a total of 9 images for comparison with the reference fraction.

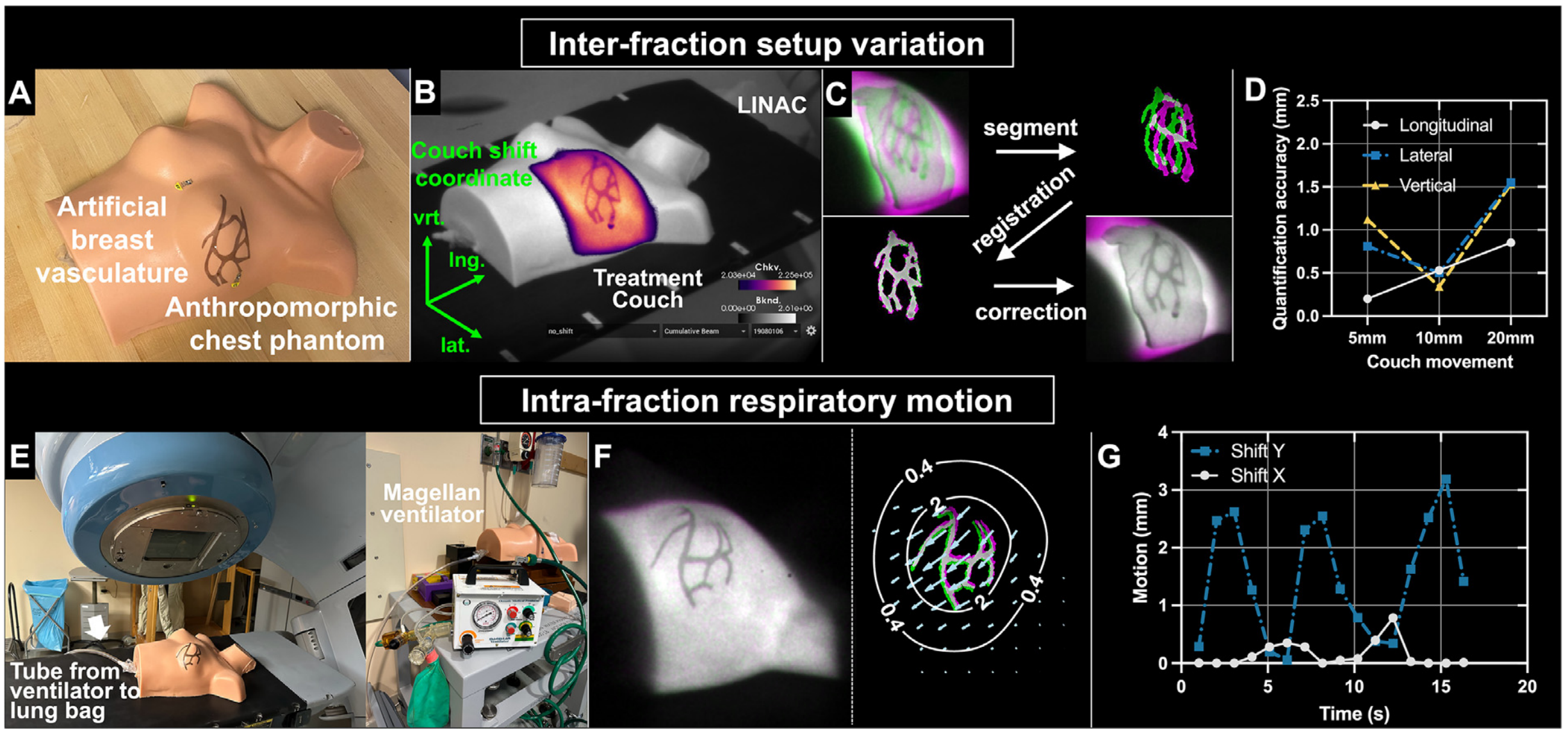

**Figure 3** Phantom experiments for testing the methodology quantification accuracy. (A-D) Test of quantifying interfraction setup uncertainties. (E-G) Test of quantifying intrafraction motions. (A) An anthropomorphic chest phantom with artificial vasculature on the breast surface. (B) Phantom setup with simulated radiation therapy treatment for quantifying interfraction setup variations that were simulated by couch movements. (C) Quantification methodology was performed on a 20 mm lateral (lat) couch movement example. (D) Quantified discrepancies for all couch movements in 3 directions: lat, longitudinal (lng), and vertical (vrt). (E) A phantom setup for monitoring intrafraction respiratory motion was provided by a Magellan ventilator. (F) An example of intrafraction motion quantification results. The left panel shows the differences between Cherenkov images at the inspiration peak and expiration base frames; the right panel shows the deformations quantified by the methodology based on the corresponding bio-morphological features in 2 frames. The recorded Cherenkov video and the quantified motion video, using the methodology based on Cherenkov imaged bio-morphological features, are provided in Video E1 and E2. (G) The quantification results illustrate that the quantified respiratory motions in the $x$ and $y$ directions of the imaging plane periodically vary with time.

The quantification methodology was then performed between the 9 Cherenkov images with varying couch movement and the Cherenkov image of the reference fraction, as shown in Fig. 3C. Rigid registration was performed in order to quantify the linear couch shift. The Cherenkov camera captures variations in the $x$ and $y$ axes of its 2D imaging plane. To translate these observations into clinically relevant movements within the patient's 3D coordinate system, a trigonometric transformation was applied. This transformation considered the geometric orientation of the camera relative to the couch in the patient coordinate system. The quantified variations in the $x$ and $y$ directions of the Cherenkov imaging plane were projected into the couch's lateral, vertical, and longitudinal shifts based on the angular relationship of the camera's imaging plane to the standard patient coordinate system used in RT. The quantified couch movement in 3 directions was then compared with the actual couch movement to test the accuracy of the methodology and its ability to determine interfraction setup errors. Zero mm of a discrepancy between quantified and actual couch movement was considered the maximal accuracy of the methodology that should ideally be reached.

#### Intrafraction motions

The same chest phantom with artificial blood vessels was set up for the quantification of intrafraction motions, as illustrated in Fig. 3E. The same plan in section 2.4.1 was deployed for the irradiation. The intrafraction respiratory motion was introduced by a Magellan pneumatically powered ventilator (Oceanic Medical Products, Inc), which connected an oxygen source valve to a cardiopulmonary resuscitation manikin lung bag inside the chest phantom through a plastic tube. During this single-fraction treatment, 20 breaths per minute with a tidal volume of 0.75 L was set for the simulated respiratory motion. Cherenkov's video of the entire treatment was recorded with a frame rate of 19.6 frames per second. In order to improve the image signal-to-noise ratio for later quantification methodology, each 20 consecutive frames was summed into a subcumulative Cherenkov image. The Cherenkov video of the intrafraction respiratory motion of the phantom is shown in Video E1.

The quantification methodology was then performed between all moving subcumulative Cherenkov images and the first reference subcumulative Cherenkov image in the video, as illustrated in Fig. 3F. Nonrigid registration was performed in order to quantify the nonlinear locoregional deformation during the intrafraction respiratory motion. A 2D map indicating the local tissue deformation because of the induced respiratory motion was generated for each moving subcumulative Cherenkov image. The quantified 2D deformation map for each subcumulative Cherenkov image is illustrated in Video E2. A deformation vector with maximal magnitude was selected as the representative one from each 2D deformation map for testing the rationality of the phantom deformation magnitude and the periodicity of respiratory motion.

### Statistical analysis

The statistical analysis in this study was performed in GraphPad Prism version 10.2.0 (GraphPad Software). A paired $t$ test evaluated statistical differences between actual and quantified variations in the phantom study on interfraction setup uncertainty. A paired $t$ test was performed to evaluate the statistical differences between the variations quantified by the rigid registration and nonrigid registration in the retrospective patient study. A $P$ value of .05 or less was considered statistically significant.

## Results

### Phantom study

The accuracy was validated to be $0.83 \pm 0.49$ (average $\pm$ 1 SD), ranging from 0.20 to 1.55 mm of discrepancy for the simulated interfraction setup uncertainties of couch translations up to 20 mm in 3 directions illustrated in section 2.4.1. The quantified couch movement was compared with actual couch movement, and the accuracy results are plotted in Fig. 3D. A paired $t$ test revealed no statistically significant difference ($P$ value = .29 > .05) between quantified and actual couch movement, indicating the accuracy of methodology on quantifying the interfraction setup uncertainties. For the test of intrafractional respiratory motion monitoring, the quantified phantom deformation magnitude in both the $x$ and $y$ direction is within the range of 8 to 10 mm maximum as measured using a physical ruler when the ventilator was on, and the quantified motion varied periodically in agreement with the respiratory pattern.

### Quantifying patient positioning accuracy

With the developed methodology, a retrospective Cherenkov imaging data set for a cohort of 10 breast cancer patients (including 5 right and 5 left breast cancer

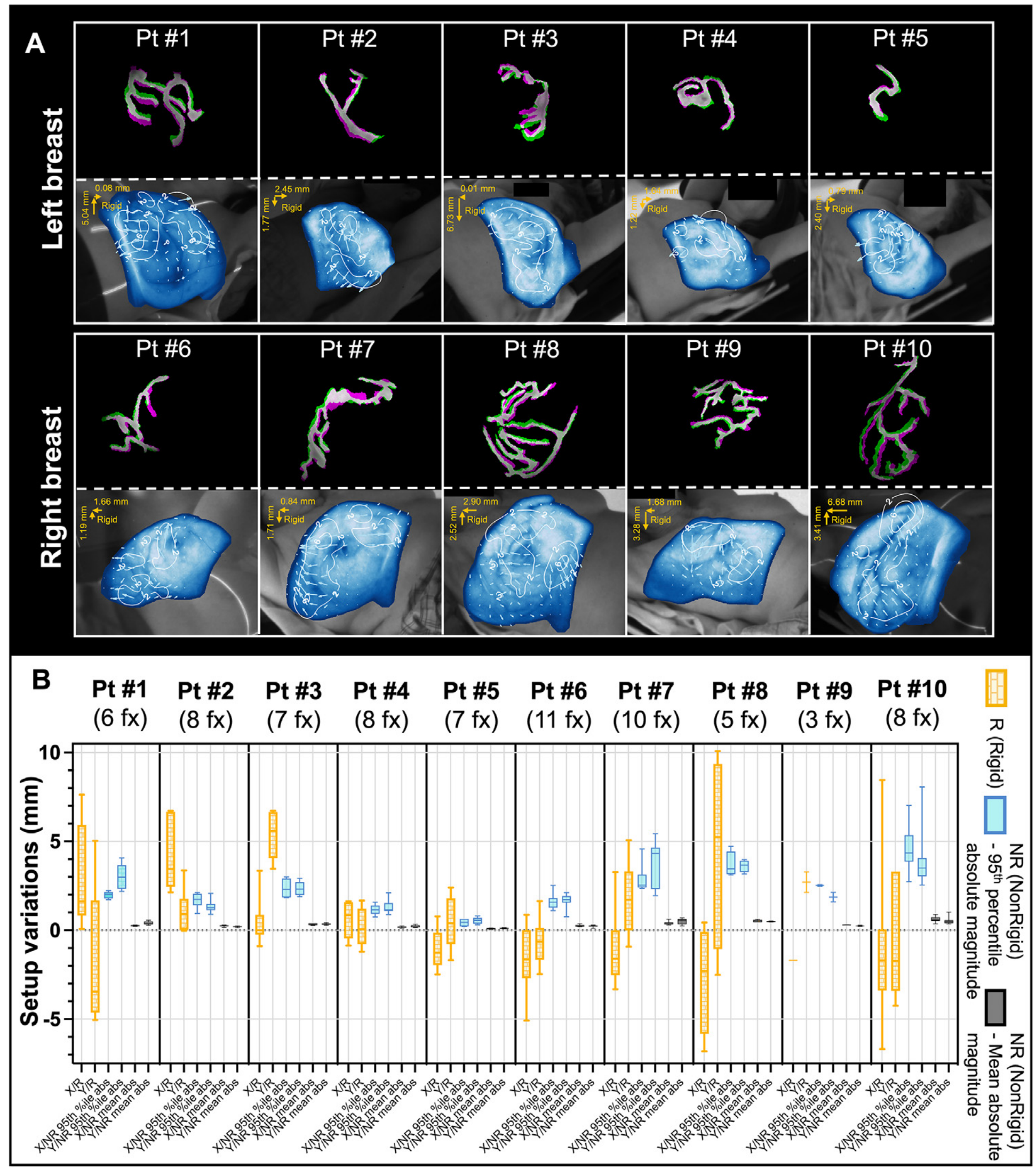

**Figure 4** Interfraction setup variations quantification results on 10 breast cancer patients (Pt; 5 left and 5 right). (A) Quantified interfraction setup variations exist between their first and second treatment fractions (fx) for 10 Pts. For each Pt, the top panel shows the variations between segmented bio-morphological features of the 2 fx. The panel underneath illustrates the quantified setup variations, including the global shifts (in orange arrows in the upper left corner) from the rigid (R) registration and the loco-regional deformation vectors (in light blue arrows overlaid on the Cherenkov image) from the nonrigid (NR) registration. The white contour is iso-curves with the same deformation magnitude. (B) Comparison of quantified setup variations: the global shifts (orange boxes) from the R registration and the loco-regional deformation (blue and black boxes) from the NR registration within all 10 Pts in the $x$ and $y$ direction of the 2-dimensional (2D) imaging plane. The blue and black boxes indicate the remaining loco-regional deformations in the quantified 2D map from the NR registration performed after the R registration, where blue boxes are the 95th percentile absolute magnitude of vectors in the 2D deformation maps while black boxes are the mean absolute magnitude of these vectors.

*Abbreviations:* X/R = *X* direction by the Rigid registration; Y/NR = *Y* direction by the Non-Rigid registration; 95th %ile abs = 95th percentile absolute magnitude; mean abs = mean absolute magnitude.

patients) was analyzed. The workflow examined the patient setup variations existing among the treatment of the 10 patients, and the results revealed a variation of 3.7 ± 2.4 mm in the magnitude of global shift. Additionally, a 2D deformation map was generated to quantify loco-regional deformation in addition to conventional global shifts between the first and subsequent fractions. Figure 4A shows the 2D deformation map for each patient between their first and second treatment fraction. Figure 4B shows the comparison of quantified variations from the rigid and nonrigid registrations.

Following the rigid registration, a nonrigid registration was then performed. Net deformation of 3.3 ± 1.9 mm was quantified as the 95th percentile of all the loco-regional vectors' magnitude in the 2D deformation maps, demonstrating that some loco-regional deformation remained after the rigid registration and then captured by the nonrigid registration. An average deformation vector of 0.12 ± 0.09 mm from all these vectors was quantified, demonstrating that loco-regional deformations are typically in various directions, so these pixel-level vectors cancel each other out after adding. Significant differences (paired $t$ test, $P$ value < .0001) in their quantified variations from 2 types of registration indicate that the rigid registration captured the majority of global variations, while the nonrigid registration quantifies the residual loco-regional deformations.

The variation found in this retrospective data set is clinically acceptable for treatment, but at present, this is a very small data set, so we do not suggest any change to clinical practice. However, this work reveals the presence of underlying setup uncertainties for the first time, and the methodology is scalable, so it can be used to analyze larger cohorts of patient data. While more clinical data are needed to quantify the exact clinical impact, and a larger follow-up study is warranted, our current findings show that the average variation of 3.7 mm minimally exceeds the 3 mm setup tolerance of the SGRT technique used in our clinic. Additionally, it is typical for clinicians to set breast planning target volume (PTV) margins to 5 to 10 mm, which may be tailored in the future based on measured variations in a larger data set using this methodology, potentially sparing more normal tissue while keeping the variations within tolerance. This work highlights the need for developing advanced patient setup and monitoring techniques, such as Cherenkov imaging, to complement current approaches and help eliminate the underlying setup uncertainties.

## Discussion

The use of Cherenkov imaged bio-morphological features in this study marks an advancement in patient positioning techniques for breast RT. By quantifying loco-regional deformations using segmented bio-morphological features like vasculature, the methodology addresses an aspect of radiation treatment precision that has been previously underexplored. For the first time, local-regional deformation is quantified based on Cherenkov imaged vasculature, opening a new avenue to improve the precision of patient positioning. The ability to detect and quantify these deformations has the potential to improve treatment protocols by allowing patient-specific adjustments or by informing the required margins needed for robust PTV definitions.

The impact of this methodology extends beyond a technical innovation, representing a step toward personalized RT treatments. By enabling the precise measurement of tissue deformation during RT, the methodology supports clinicians in making informed decisions regarding dosimetric and position adjustments when necessary, as well as rational design of margins to define the PTV. This capability is particularly vital in treatments where millimeter-level precision can significantly influence both the efficacy of tumor targeting and the preservation of surrounding healthy tissues.[20,21] Furthermore, the bio-morphological features detected through Cherenkov imaging are unique to each individual, which could be applied to verify patient identities through the treatment courses, adding safety redundancy.

Admittedly, there are limitations related to the dimensionality and sensitivity in its current stage. Because of Cherenkov imaging capturing 3D objects within 2D imaging space, the current methodology in the phantom study showed heightened sensitivity to variations along vertical and longitudinal directions compared with lateral movements. This sensitivity bias in different dimensions is because of the use of only 1 camera. A potential solution to overcome this limitation would be using 2 Cherenkov cameras to reconstruct the 3D patient surfaces with stereovision algorithms, allowing for more accurate spatial quantification of deformations.[22,23] Other work has demonstrated the feasibility of quantifying the 3D surface deformation by deformable registration based on SGRT systems.[24]

The SGRT setup tolerances for breast external beam radiation therapy prior to treatment in this study are ≤3 mm in the isocenter position and ≤3 degrees in gantry angle.[25,26] Our analysis shows that 54.0% (34 out of 63 fractions) of later treatments after the first treatment fraction among the 10 breast cancer patients are outside this preset and monitored SGRT tolerance (3 mm and 3 degrees). This would be because of several reasons: first, the translational and rotational discrepancies are combined in the 2D Cherenkov image domain, leading to larger apparent translational discrepancies; second, the effect of intrafractional motion such as respiratory motion is integrated into the cumulative Cherenkov images, contributing to larger discrepancies. The quantified setup variations of 3.7 ± 2.4 mm in global shift and 3.3 ± 1.9 mm in loco-regional deformation indicate the potential value of the Cherenkov methodology in this work to be added on SGRT for more accurate patient setup. These reconstruction results could potentially be integrated into currently deployed SGRT systems, which is a work in progress in collaboration with SGRT industrial partners.[26] An intrinsic limitation to this technique is that Cherenkov photons are only detected to a depth of up to 10 mm, meaning that deformations are observed in superficial tissue layers. It is unclear what the relationship is to deeper lying tissues, especially in larger patients.

Addressing loco-regional deformations remains a complex challenge. The clinical impact of these deformations remains unclear in various scenarios, but the quantification is a crucial step forward for prospective studies evaluating the dosimetric effects. Currently, the segmentation of bio-morphological features is semimanual, which is not only subjective but also inefficient. A deep-learning approach (eg, U-Net) is being developed to automate this process, which could significantly enhance the accuracy and speed of vasculature segmentation, enabling segmentation in real-time at the point of image acquisition.[27,28] Moreover, the applicability of our method is limited to scenarios where bio-morphological features are clearly visible in Cherenkov images. This visibility is often compromised in highly modulated dynamic treatments such as volumetric modulated arc therapy, suggesting a need for further technological advancements to adapt this method for broader clinical use.

## Conclusions

This study introduced a novel application of Cherenkov imaging, using the registration of bio-morphological features to precisely quantify loco-regional tissue deformation. These directly observed measures from both phantom and human images showed the method's capacity to measure both global and local variations in patient positioning with a high degree of precision. Both rigid and nonrigid registrations of Cherenkov imaged vasculature provide robust methods for quantifying the patient positioning variations and potential inter- and intrafraction adjustments. Quantified setup variations with the level of 3.7 ± 2.4 mm of global shift and 3.3 ± 1.9 mm of loco-regional deformation at the 95th percentile of the vector magnitude were observed existing among 10 breast cancer patients but being ignored during their treatment courses. This approach is especially useful for breast cancer RT, given the high amount of soft tissue local deformation that can occur in daily positioning variation. Work is ongoing to integrate this methodology into clinical workflows to provide corrective feedback. This kind of methodology provides one possible path to personalized, precise RT treatments, optimizing tumor targeting while minimizing exposure to surrounding healthy tissues.

## Disclosures

The authors Savannah M. Decker, Petr Bruza, Lesley A. Jarvis, and Brian W. Pogue are affiliated with DoseOptics, LLC, who provided hardware support for this study.

## Acknowledgments

The authors are grateful to Dr Venkat Krishnaswamy for providing the equipment and support for the respiratory motion phantom study. Yao Chen was responsible for statistical analysis.

## Supplementary materials

Supplementary material associated with this article can be found in the online version at doi:10.1016/j.adro.2024.101684.